\documentclass[a4paper,useAMS,usenatbib,onecolumn]{mnrasfile}
\usepackage{amsmath}
\usepackage{hyperref}
\usepackage{amssymb}
\usepackage{graphicx}
\usepackage{enumerate}
\usepackage{subfigure}
\usepackage[T1]{fontenc}
\usepackage{aecompl}
\usepackage{verbatim}
\usepackage{float}
\usepackage{lipsum}
\usepackage{enumerate}

\title[Supervised estimator for matter power spectrum -- {\rm SEMPS}]{A supervised machine learning estimator for the non-linear matter power
spectrum -- SEMPS}

\author[Mohammed $\&$ Verma] {Irshad
  Mohammed,\thanks{irshad@physik.uzh.ch}$^{1,2}$ and 
  Janu Verma \thanks{j.verma5@gmail.com}$^3$\\ \\
  $^1${Physik-Institut, University of Zurich,
  			Winterthurerstrasse 190, 8057 Zurich, Switzerland} \\
  $^2${Institute for Computational Science, University of Zurich,
  			Winterthurerstrasse 190, 8057 Zurich, Switzerland} \\
  $^3${Institute for Genomic Diversity, Cornell University, Ithaca 14853, USA}
}
\begin{document}
\maketitle
\begin{abstract}
	In this article, we argue that models based on machine learning (ML) can be
    very effective in estimating the non-linear matter power spectrum ($P(k)$).
    We employ the prediction ability of the supervised ML algorithms to build
    an estimator for the $P(k)$. The estimator is trained on a set of
    cosmological models, and redshifts for which the $P(k)$ is known, and it
    learns to predict $P(k)$ for any other set.
	We review three ML algorithms --- Random Forest, Gradient Boosting Machines,
    and K-Nearest Neighbours --- and investigate their prime parameters to
    optimize the prediction accuracy of the estimator. We also compute an
    optimal size of the training set, which is realistic enough, and still yields
    high accuracy.
	We find that, employing the optimal values of the internal parameters,
    a set of $50-100$ cosmological models is enough to train the estimator
    that can predict the $P(k)$ for a wide range of cosmological models, and
    redshifts.
    Using this configuration, we build a blackbox --- Supervised Estimator for
    Matter Power Spectrum (SEMPS) --- that computes the $P(k)$ to 5-10$\%$
    accuracy up to $k\sim 10 h^{-1}{\rm Mpc}$ with respect to the reference
    model (cosmic emulator). We also compare the estimations of SEMPS to that of
    the Halofit, and find that for the $k$-range where the cosmic variance is
    low, SEMPS estimates are better than that of the Halofit.
	The predictions of the SEMPS are instantaneous in the sense that it can
    evaluate up to 500 $P(k)$ in less than one second, which makes it ideal for many
    applications like visualisations, weak lensing, emulations, likelihood
    analysis etc..
	As a supplement to this article, we provide a publicly available software
    package\footnote{\url{http://www.ics.uzh.ch/~irshad/semps/}}.


\end{abstract}

\begin{keywords}

(cosmology:) cosmological parameters,
cosmology: miscellaneous,
(cosmology:) large-scale structure of Universe,
methods: numerical,
methods: statistical.

\end{keywords}
\section{Introduction}\label{sec:intro}

Recently the Planck satellite, employing the cosmological information embedded
in the anisotropies of the cosmic microwave background, has put the tightest
constraints on the cosmological parameters \citep{2015arXiv150201589P}. This
marks the advent of the era of precision cosmology. However, there is a wealth
of cosmological information in the large scale structures of the Universe which
is not exploited by Planck. The future generation surveys like Euclid
\citep{2013LRR....16....6A} and LSST \citep{2009arXiv0912.0201L} are expected to
employ this information to take precision cosmology to the next level.

To achieve this, a thorough understanding of the non-linear structure formation
and growth is vital. As the large scale structures of the Universe are the
result of gravitational instability of the tiny initial matter density
fluctuations, which is a random field, no theory can predict the distribution of
matter at any epoch in the cosmic history of time. The best one can hope to
achieve is to predict the statistical properties of the density field. The
simplest of which is the two-point function or the power spectrum of the density
field. In order to employ the constraining power of the future surveys, an
accurate modelling of the matter power spectrum is needed up to highly
non-linear regime $k\sim 10\  h^{-1}\mathrm{Mpc}$.

There are various methods to estimate the matter power spectrum which have been successfull at various scales. Perturbation theories (for a review see
\cite{1994FCPh...15..209D,2002PhR...367....1B}) give the most accurate results,
but only up to $k\sim 0.2\ h^{-1}\mathrm{Mpc}$. The halo model
\citep{1977ApJ...217..331M,2000MNRAS.318..203S,2002PhR...372....1C} framework is
computationally faster, but its accuracy is limited only up to $\sim 30\%$ in
the non-linear regime.  However, some recent techniques, employing modified
version of the halo model, provide very accurate results
\citep{2014MNRAS.445.3382M,2015PhRvD..91l3516S,2015arXiv150507833M}.
Halofit \citep{Smith:2002dz,2012ApJ...761..152T} provides fitting functions based
on cosmological simulations, and the matter power spectrum so evaluated are
accurate up to 10$\%$ in the non-linear regime. An implementation of the Halofit,
called CAMB \citep{Lewis:1999bs,Lewis:2007kz,Challinor:2011bk} is widely used.
Numerical simulations are the ultimate solution. \cite{2010ApJ...715..104H} has
developed a cosmic emulator (henceforth Emulator) to evaluate the matter power
spectrum based on 38 $N$-body cosmological simulations. Emulator power spectra
are accurate to 3-5 percent up to $k\sim 10\  h^{-1}\mathrm{Mpc}$ on their
original cosmological nodes, and lesser so on other sets of cosmological
parameters.  In addition to the accuracy, the speed of the computation is also an
important factor. For example, if the computation of matter power spectrum takes
few seconds, the projected weak lensing shear power spectrum will take few
minutes, and thus the likelihood analysis of the weak lensing data becomes
highly expensive. The developments in this subject in the last few years have
been very impressive.

Machine learning has proven to be very effective in solving complex problems in technology (e.g. search, recommendation, spam detection), computational genomic (\cite{Larrañaga01032006}, \cite{Jensen15122011}), healthcare (\cite{aba}, \cite{abcd}), finance \citep{2011arXiv1107.0036B} etc. There has been considerable progress in bridging the gap between machine learning and computational astrophysics, for example the development of the AstroML library (\cite{astroML}, \cite{astroMLText}). In this paper, we employ machine learning based models to estimate the matter
power spectrum with great accuracy and speed.
We apply the framework of supervised machine learning where a model is trained
on a set of cosmological simulations, and it learns to predict the non-linear
matter power spectrum over a range of cosmological models, and redshifts with a
few percent accuracy. Though the algorithm takes a few seconds to train, the
predictions are instantaneous which are ideal for likelihood analyses, e.g.,
MCMC. We provide a software package (SEMPS) which contains the model trained on a set of
cosmologies and redshits. The power spectrum values can be called from SEMPS for new cosmologies and redshifts. Calls to the trained model
retrieve almost instantaneous results for a range of $k-$values. The accuracy of such a
model will increase as more relevant data (e.g. from independent simulations) is included in the training set.

The advantage over simulations and semi-analytic models is its (i) simplicity,
(ii) accuracy which improves with the availability of new data, and
(ii) instantaneous speed of predictions.

This paper is organised as follows: We start by introducing the basics of
machine learning, and explain the algorithms used in section \ref{sec:ml}.
In section \ref{sec:data}, we introduce various datasets, and their configurations
important to the analysis. In section \ref{sec:params} we perform an analysis of
the intrinsic parameters of the algorithms, and requirements for the data.
In section \ref{sec:predictions}, we analyse the performance and accuracy of
the predictions for all datasets for their role as training and test set.
In section \ref{sec:semps}, we introduce SEMPS, its main features, accuracy,
and efficiency. Finally in section \ref{sec:discussion}, we summarise the
article, discuss some important aspects, and finish with some future prospects.


\section{Machine Learning}\label{sec:ml}
\subsection{Introduction}
Machine learning refers to any systematic way of finding patterns in data, usually for the purpose of making predictions. It works by developing algorithmic models which can be programmed into computers to automatically learn the irregularities in data, and make predictions. The learning algorithm gives computers the ability to learn without being explicitly programmed. The hope is that, with suitable theoretical machinery, we can program computers to emulate the learning mechanism of the human brain. Needless to say, we are far from this goal. See \citep{5362936} for a quick introduction to machine learning.\\
\\
In this article, we will restrict to a special class of learning models which fall under {\em supervised machine learning} (chapter 1 of \cite{hastie01statisticallearning}). In these models, the learning is guided by observations with known outcomes. If we have a quantitative output, that we want to predict based on a set of {\em features}. We prepare a {\em training set}, which comprises of features and corresponding output values for a set of data points. We choose a supervised machine learning algorithm, which {\em trains} on the training set, and builds a predictive model, which in turn can predict the output for new unseen data points. \\
A data point, with its features, is called a {\em sample} and the corresponding output value is called {\em target} in machine learning lingo. \\
\\
The basic setting for a supervised machine learning model is as follows:  
\begin{itemize}
\item We have a labelled training set i.e. samples with known values of target.
\item We are given an unlabelled testing set i.e., samples for which the target values are unknown.
\item Goal is to build a model which trains on the labelled data to predict the target for the unlabelled data. 
\end{itemize}
For an introduction to basics of machine learning, refer to the excellent text by Tom Mitchell \citep{Mitchell:1997:ML:541177}. A more mathematical treatment of the subject can be found in \citep{hastie01statisticallearning}.
In this article, we will describe three machine learning algorithms for predicting the matter power spectrum.

\subsection{Algorithms}\label{sec:algorithms}
For the current study, we employed the following supervised machine learning algorithms which will be explained further.
\begin{itemize}
\item Random Forests (RF)
\item Gradient Boosting Machines (GBM)
\item K-Nearest Neighbours (KNN)
\end{itemize}

In all that follows, the features are expected to be elements of a vector space of dimension equal to number of features ($p$, e.g. a set of cosmological parameters). i.e a sample is a tuple of vectors $({\mathbf x},y)$, where
\begin{equation}
{\mathbf x} = (x_1, x_2, \ldots, x_p) \in \mathbb{R}^{p}
\end{equation}
and $y \in \mathbb{R}$ is the target value. \\
Before we delve into {\em ensemble methods} (RF, GBM), we have to understand how {\em decision tree} works for regression.\\
\\
{\bf Decision Trees :}\\
Decision trees \citep{ig}, \citep{Quinlan:1993:CPM:152181} are binary trees, which are formed by recursively partitioning the feature space into smaller, manageable (homogeneous) chunks. The algorithm works by fitting simple models to these chunks. Thus, the complete model comprises of two parts - recursive partitioning of the feature space to construct a tree, and fitting simpler models to the smaller subspaces created in the first step.\\
\\
For example, consider the situation with only 2 features i.e. 
${\mathbf x} = (x_1,x_2)$ and the output $y \in \mathbb{R}$. Let the data points be 

\begin{equation}
\begin{array}{l}
\displaystyle {\mathbf x} = (0.2, 1.3), y = 0.4 \\
\displaystyle {\mathbf x} = (0.4, 1.5), y = 0.5 \\
\displaystyle {\mathbf x} = (1.2, 2.1), y = 0.9 \\
\displaystyle {\mathbf x} = (1.1, 2.4), y = 1.0
\end{array} 
\label{eqn:1}
\end{equation}

This data can be partitioned into two subsets as follows :
\begin{equation}
H_1 = (x_1 < 1, x_2 < 2)  \hspace{2mm} \text{and} \hspace{2mm} H_2 = (x_1 > 1, x_2 > 2)
\end{equation}
Note that this is one of the possible trees, the power of algorithm comes from the fact that it picks best possible tree, in the sense which will be made clear soon. \\
\\
The algorithm automatically decides the variables to split, value of the variable where split occurs, and the topology of the tree. How does it do that ? \\
\\
The tree building proceeds with a greedy algorithm. Let $x_i$ be the splitting variable, with split point $x_i = l$, then we get a partition of the feature space as - 
\begin{equation}
H_1 = \{ {\mathbf x} | x_i < l \} \hspace{2mm} \text{and} \hspace{2mm} H_2 = \{ {\mathbf x} | x_i > l \} 
\end{equation}
And assign target values to each compartment as 
\begin{equation}
\hat{y} = 
\begin{cases}
c_1, & \text{for ${\mathbf x} \in H_1$} \\
c_2, & \text{for ${\mathbf x} \in H_2$}
\end{cases}
\end{equation}
The desired splitting should minimize the total {\em sum of squares error} in each compartment of the partition of the feature space. 
\begin{equation}
S = \sum_{H_i} \sum_{x_j \in H_i} (y_j - c_j)^2 
\label{eqn:S}
\end{equation}
It is not hard to check that the minima occurs at 
\begin{equation}
\hat{c_j} = {\rm Mean} \{y_j | x_j \in H_i \}
\end{equation}
This means that each element in the subspace $H_i \subset \mathbb{R}^p$ is assigned a target value which is equal to the average of the target values in $H_i$. This is equivalent to fitting a linear regression in each of the subspaces of the feature space.\\
Notice, we are partitioning, and fitting a simpler model to the training set simultaneously. \\
\\
In the above example (equation \ref{eqn:1}), this translates to the assignment 
\begin{equation}
\hat{y} = 
\begin{cases}
0.45, & \text{for $x \in H_1$} \\
0.95, & \text{for $x \in H_2$}
\end{cases}
\end{equation}
To summarize, the regression-tree algorithm is as follows : 
\begin{enumerate}
	\item Start with a single node containing all the points, no partition.
	\item Calculate the estimated target value, and sum of squares error, $S$
	\item Search over all binary splits of all variables to find one which minimizes $S$. 
	\item In each new node, repeat from step (i). 
\end{enumerate}
These tree-based methods are very simple conceptually, and are easy to program, yet very potent. We used two such methods in this work, which are described in subsequent sections. \\
More details on decision trees can be found on chapter 3 of \citep{Mitchell:1997:ML:541177}.

\subsubsection{Random Forests {\rm (RF)}}\label{subsec:random forests}

Random forests \citep{Breiman:2001:RF:570181.570182} is one of the most powerful machine learning algorithms. A random forest is built from an {\em ensemble} of decision trees. The idea of ensembling methods (chapter 16 of \cite{hastie01statisticallearning}) is to build several models independently, and then the output is taken to be the average of the outputs from the models in the ensemble. \\
\\
In random forests, an ensemble of trees is constructed where each tree is constructed on a random sample drawn from the training set. The sample is drawn with replacement, it is called {\em bootstrap sampling} \citep{efron1979}, \citep{tEFR93a}. \\
The idea behind RF is to generate multiple little trees from random subsets of data. In that way, each of those small trees gives some group of ill-conditioned (biased) estimators (chapter 15 of \citep{hastie01statisticallearning}). Each of them is capturing different regularities, since random subset of the instances are in the interest. At the extreme randomness, it curates nodes from random subset of the features as well. In this way feature based randomness is also used. After you simply create $m$ number of trees in this random way, we are able to obtain more cluttered decision boundaries than the simple lines. See chapter 15 of \citep{hastie01statisticallearning} for more details.

\subsubsection{Gradient Boosting Machines {\rm (GBM)}}\label{subsec:gbm}

Gradient boosting machines \citep{friedman2001}, \citep{Friedman:2002:SGB:635939.635941} are a class of very effective machine learning algorithms that have shown considerable success in a wide variety of applications. These methods are based on the concept of {\em boosting} \citep{Breiman96bias}, \citep{Freund1997119}, which means a way to convert {\em weak learners} to {\em strong} ones. It provides an answer to the question \citep{Kearns:1989:CLL:73007.73049} : \\
\begin{center}
{\em Can a set of weak learning algorithms build a strong one ?}. \\
\end{center}
Let our training set be 
\begin{equation}
T = \{ ({\mathbf x}_1,y_1), ({\mathbf x}_2,y_2), \ldots, ({\mathbf x}_p, y_p) \}
\end{equation}
where each ${\mathbf x}_i \in \mathbb{R}^N$ and each $y_i \in \mathbb{R}$. Both ${\mathbf x}_i$ and $y_i$ values are known for the training examples. \\
The goal of supervised learning is to approximate a {\em decision function} $F({\mathbf x})$, and the estimated value $\hat{y} = F({\mathbf x})$, such that an {\em error function} ($E(y, F({\mathbf x}))$) is minimized. In case of linear model, the error function is chosen to be the sum of squares error. This is the same error function, we used for decision trees as well (equation \ref{eqn:S}). \\
GBM assumes that the decision function is a linear combination of decision functions of a set of weak learning models. i.e. 
\begin{equation}
F({\mathbf x}) = \sum_{i=1}^{K} \gamma_{i} h_{i}({\mathbf x})
\end{equation}
where $h_i$'s are the decision functions of the weak learners and $\gamma_i$'s are the corresponidng linear coefficients.
In this study, we take the weak learners to be the decision trees. Decision trees are particularly valuable for boosting, as they can handle data of mixed type, and model complex functions. In the following, each $h_i({\mathbf x})$ is the output of a decision tree trainer.\\ 
If $E(y, F({\mathbf x}))$ is the error function, we want to find its minima. The GBM algorithm proceeds via gradient descent iteratively as follows : 
\begin{enumerate}
\item Start with a model where each sample is given a constant target value. 
\begin{equation}
F_0({\mathbf x}) = \text{arg min}_{\gamma} \left( \sum_{j=1}^N E(y_j, \gamma) \right)
\end{equation}
Here $\text{arg min}_{x}(f(x))$ refers to the value of the argument ($x$) such that the function ($f$) attains its minimum value. In the above situation, the left side of the equation computes the value of $\gamma$ for which the sum term is minimized. 
\item For $i=1$ to $K$ :
\begin{enumerate}
\item Compute the {\em residuals}
\begin{equation}
r_{ij} = \frac{\partial E(y_j, F_{i-1}({\mathbf x}_j))}{\partial F_{i-1({\mathbf x}_j)}}
\end{equation}
\item Compute the gradient jump
\begin{equation}
\gamma_i = \text{arg min}_{\gamma} \left( \sum_{j=1}^N E(y_j, F_{i-1}({\mathbf x}_j) - \gamma r_{ij}) \right) 
\end{equation}
\item Update the decision function
\begin{equation}
F_i({\mathbf x}) = F_{i-1}({\mathbf x}) + \gamma_i h_i({\mathbf x}) 
\end{equation}
\end{enumerate}
\item Compute $F_K({\mathbf x})$. 
\end{enumerate} 
Since we are taking derivatives of the error function in this procedure, it is evident that this algorithm would work for any differentiable error function. For more details refer to chapter 10 of \cite{hastie01statisticallearning}


\subsubsection{K-Nearest Neighbours {\rm (KNN)}}\label{subsec:knn}

K-Nearest neighbours \citep{doi:10.1080/00031305.1992.10475879} is one of the simplest machine learning algorithms which falls under {\em instance-based learning} methods (chapter 8 of \cite{Mitchell:1997:ML:541177}), which means that the computation is deferred until a test case is supplied.\\
The KNN algorithm works by finding K samples from the training set which are closest to the test sample, and the test sample is assigned the target value which is equal to the mean of the target values in the chosen neighbourhood. \\
\\
As earlier, the feature space of the samples is a real-vector space, and each feature vector has a corresponding target label. To compute the degree of closeness of samples, some distance metric is employed. \\
The algorithm can be described schematically as follows : 
Let our training set be 
\begin{equation}
T = \{ ({\mathbf x}_1,y_1), ({\mathbf x}_2,y_2), \ldots, ({\mathbf x}_N, y_N) \}
\end{equation}
where each ${\mathbf x}_i \in \mathbb{R}^p$ and each $y_i \in \mathbb{R}$. Both ${\mathbf x}_i$ and $y_i$ values are known for the training examples.\\
Let ${\mathbf x}^{\prime}$ be a test instance, whose target value we want to predict. 
\begin{enumerate}
\item For each ${\mathbf x}_i$ in training set, compute its distance from the test case ${\mathbf x}^{\prime}$. 
\begin{equation}
{\mathfrak D} = \{ ({\mathbf x}, y, d({\mathbf x},{\mathbf x}^{\prime})) | ({\mathbf x},y) \in T \}
\end{equation}
where $d({\mathbf x},{\mathbf x}^{\prime})$ is the distance between ${\mathbf x}$ and ${\mathbf x}^{\prime}$. 
\item Sort ${\mathfrak D}$ in increasing values of $d$. 
\item Find the set of K training objects which are closest to $({\mathbf x}^{\prime})$ i.e. 
\begin{equation}
S_{\rm K} \equiv \text{first K entries of} \hspace{1mm} {\mathfrak D}
\end{equation}
\item Compute the predicted target value for ${\mathbf x}$.
\begin{equation}
\hat{y} = \frac{1}{\rm K} \sum_{i=1}^{\rm K} y_i
\end{equation}
\end{enumerate}
There are various choices for distance metric e.g Euclidean, Manhattan, Cosine similarity, Minkowski etc. \\
Another important parameter of this algorithm is the choice of K, number of neighbours are used to computed the prediction. We will study the imortance of K in later sections.\\
\\
KNN is very fast, its speed is proportional to the number of samples in the training data. Its simplicity and speed makes it an ideal model for comparison.

\section{Datasets}\label{sec:data}

\begin{table}
	\centering
	\begin{tabular}{ l | c | l }
	  \hline
	  Dataset & Number of cosmological models & Remarks\\
	  \hline
	  E38 & 38 & Emulator\\
	  \hline
	  ER50 & 50 &Emulator+Random\\
	  ER100 & 100 & Emulator+Random\\
	  ER1000 & 1000 & Emulator+Random\\
	  \hline

	  R50-1 & 50 & Random\\
	  R50-2 & 50 & Random\\
	  R100-1 & 100 & Random\\
	  R100-2 & 100 & Random\\
	  R1000-1 & 1000 & Random\\
	  R1000-2 & 1000 & Random\\
	  R1000-3 & 1000 & Random\\
	  \hline

	  NR100-1 & 100 & Normal\\
	  NR100-2 & 100 & Normal\\
	  NR100-3 & 100 & Normal\\
	  \hline
	\end{tabular}
	\caption{Listing all datasets.}
	\label{tbl:data}
\end{table}

\begin{table}
	\centering
	\begin{tabular}{ c | l | c | c }
	  \hline
	  Parameter & Description & Range for Uniform distribution & Mean and std.dev. for normal distribution\\
	  \hline
	  $h$ & Hubble constant in units 100 $\mathrm{km\ s^{-1}\ Mpc^{-1}}$ &
			0.65--0.75 & (0.7,0.007) \\
	  $\Omega_m$ & Normalised matter density at redshift zero
	  				 & $0.121<\Omega_mh^2<0.154$ & (0.1375/$h^2$,0.001/$h^2$) \\
	  $\Omega_b$ & Normalised baryon density at redshift zero
	  			& $0.0219<\Omega_bh^2<0.0231$ & (0.0225/$h^2$,0.0001/$h^2$) \\
	  $\sigma_8$ & Normalisation of matter power spectrum & 0.75--0.84 & (0.8,0.01) \\
	  $n_s$ & Spectral index & 0.9--1.0 & (0.96,0.01) \\
	  $w$ & Equation of state of dark-energy & $-1.1\ -\ -0.9$ & (-1.0,0.03) \\
	  \hline
	\end{tabular}
	\caption{Cosmological parameters.}
	\label{tbl:cosmo}
\end{table}

We used {\em cosmic emulator}
\citep{2010ApJ...715..104H,2009ApJ...705..156H,2010ApJ...713.1322L,2014ApJ...780..111H} to generate different datasets
containing the matter power spectrum for different cosmological models and redshifts. Table
\ref{tbl:data} shows a list of all datasets, and their particulars. We used
total 14 datasets of different sizes and configurations.

Two very important ingredients are the choice of redshift and $k$ arrays (henceforth ${\mathbf z}$ and ${\mathbf k}$ respectively).
It is desirable that the algorithm learns from various combinations
of cosmological models, ${\mathbf z}$, and ${\mathbf k}$, therefore it is more useful
to have randomised arrays in the training dataset rather than fixed. To randomise ${\mathbf k}$,
and ${\mathbf z}$, we follow these two steps:

\begin{itemize}
	\item Create an array of size $N_p$, let us call it {\it parent array},
	\item draw a random subset of size $N$ from the parent array, let us call it
			{\it data array}.
\end{itemize}

The parent array for $k$ is taken to be of size 100, which comprises : (i) 10 entries
between 0.001 and
0.025, (ii) 50 entries between 0.025 to 1, and (iii) 40 entries between 1 and 10.
In each of these ranges the numbers are equally spaced in log$_{10}$.
For each cosmological model in
every dataset, a data array of size 50 was randomly
sampled from the parent array, and power spectrum was computed using
the cosmic emulator.
The parent array for redshift is also taken to be of size 100. The entries are numbers, equally spaced between 0 and
4.0. The data array is chosen to be of size 20.
Therefore, for each cosmological model in every dataset, power spectrum
is evaluated for a total of 1000 values (50 $k\ \times$ 20 $z$).
The full dataset has a structure
\begin{equation}
	{\rm Dataset} \equiv \left\{ {\mathbb C},{\mathbf k},{\mathbf z}\right\}.
\end{equation}
\\
where, ${\mathbb C}$ is a set containing six cosmological parameters
\begin{equation}
	{\mathbb C} \equiv \left\{ \Omega_m,\sigma_8,h,n_s,\Omega_b,w \right\}.
\end{equation}
\\
Table \ref{tbl:cosmo} gives a brief description of each cosmological parameter.

We generated four categories of datasets:
\begin{itemize}
\item {\bf E38} category contains
38 cosmological models which are the original nodes of the cosmic emulator.
\item  {\bf RXXX-Y} category contains XXX sets of random cosmological parameters drawn from
a uniform distribution. The range of the uniform distribution is given in the
table \ref{tbl:data}. Y is the unique identifier for each RXXX dataset, e.g. the datasets R50-1 and R50-2 are two different datasets, each containing 50 sets of random cosmological parameters. 7 such datasets are employed in our study, 2 of size 50, 2 of size 100, and 3 of size 1000.
\item {\bf ERXXX} category contains XXX sets of cosmological parameters, 38 of which are the cosmologies from cosmic emulator
nodes, the remaining ones are drawn from a uniform distribution of the cosmological parameters (similar to RXXX sets).
For example ER100 contains E38 plus R62.
\item {\bf NRXXX-Y} category contains XXX sets of cosmological parameters, each of which is drawn form
a normal distribution. The mean and standard deviation for each parameter is given in table
\ref{tbl:data}. The unique identifier Y, in this case, resembles the width of the distribution.
For example, NR100-2 contains 100 sets of cosmological parameters randomly drawn from a
normal distribution with mean and two times the standard deviation width as listed in table \ref{tbl:data} .
\end{itemize}
\begin{figure}
	\centering
	\includegraphics[width=0.45\textwidth]
		{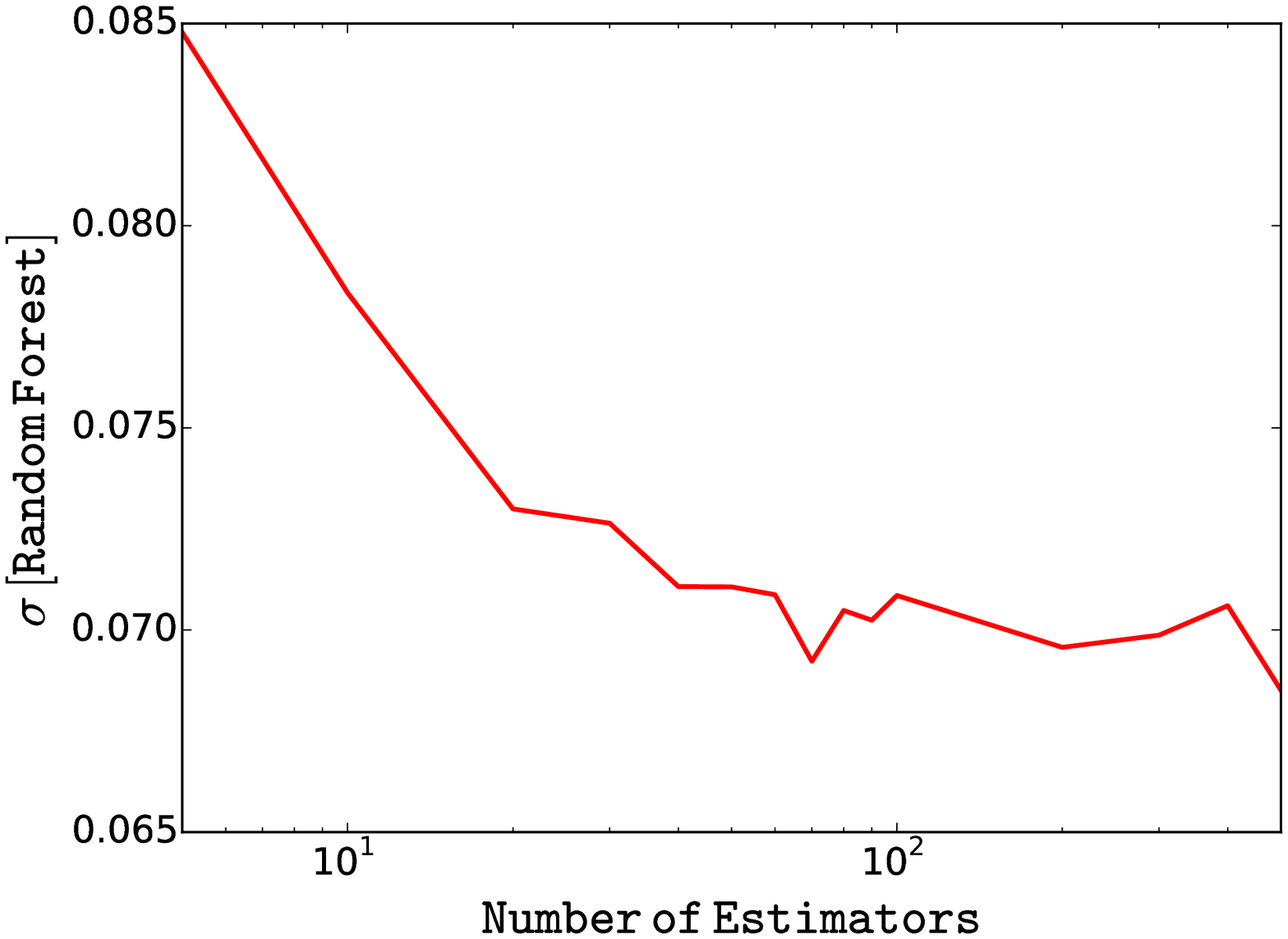}
	\includegraphics[width=0.45\textwidth]
		{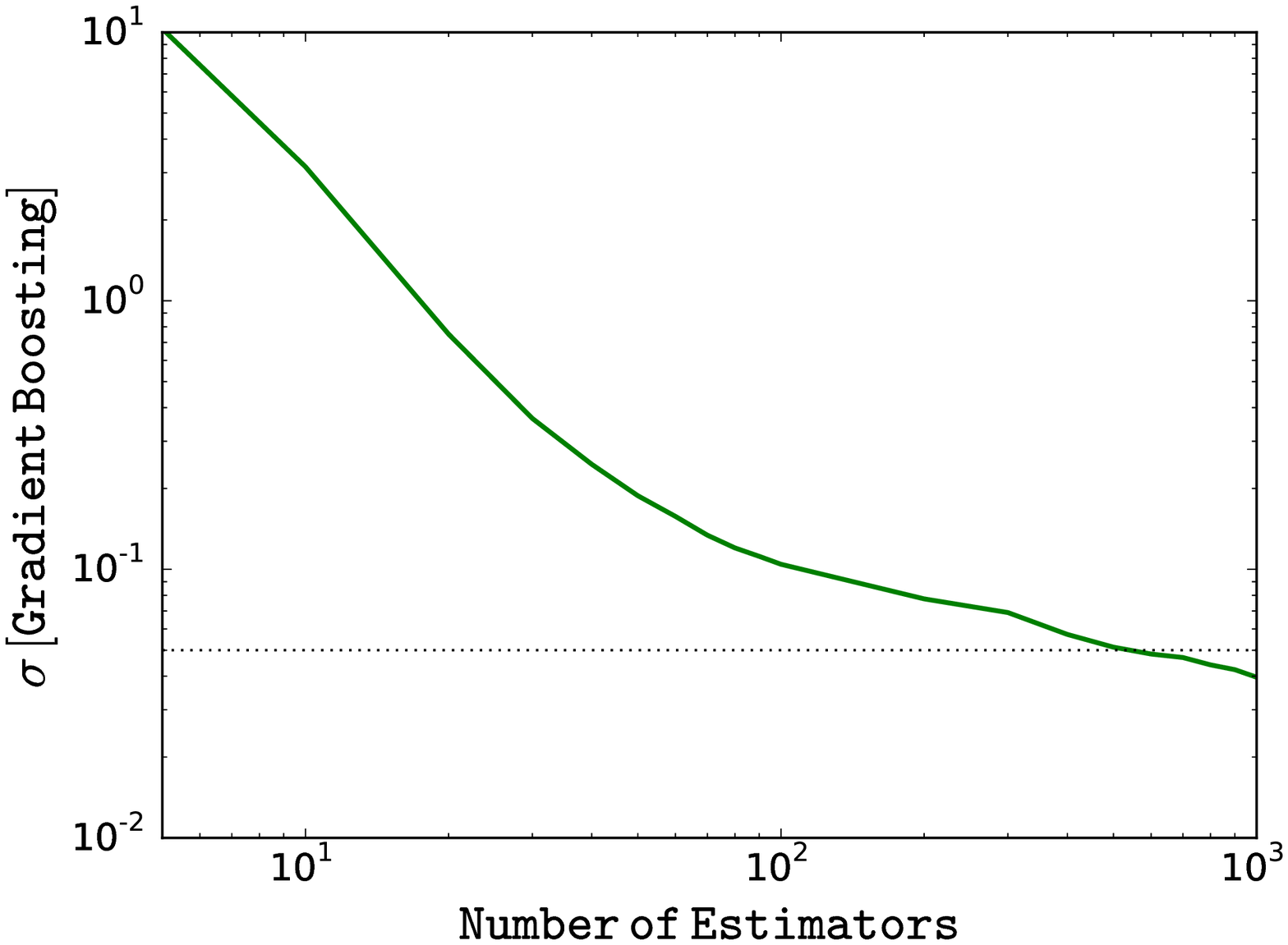}
	\includegraphics[width=0.45\textwidth]
		{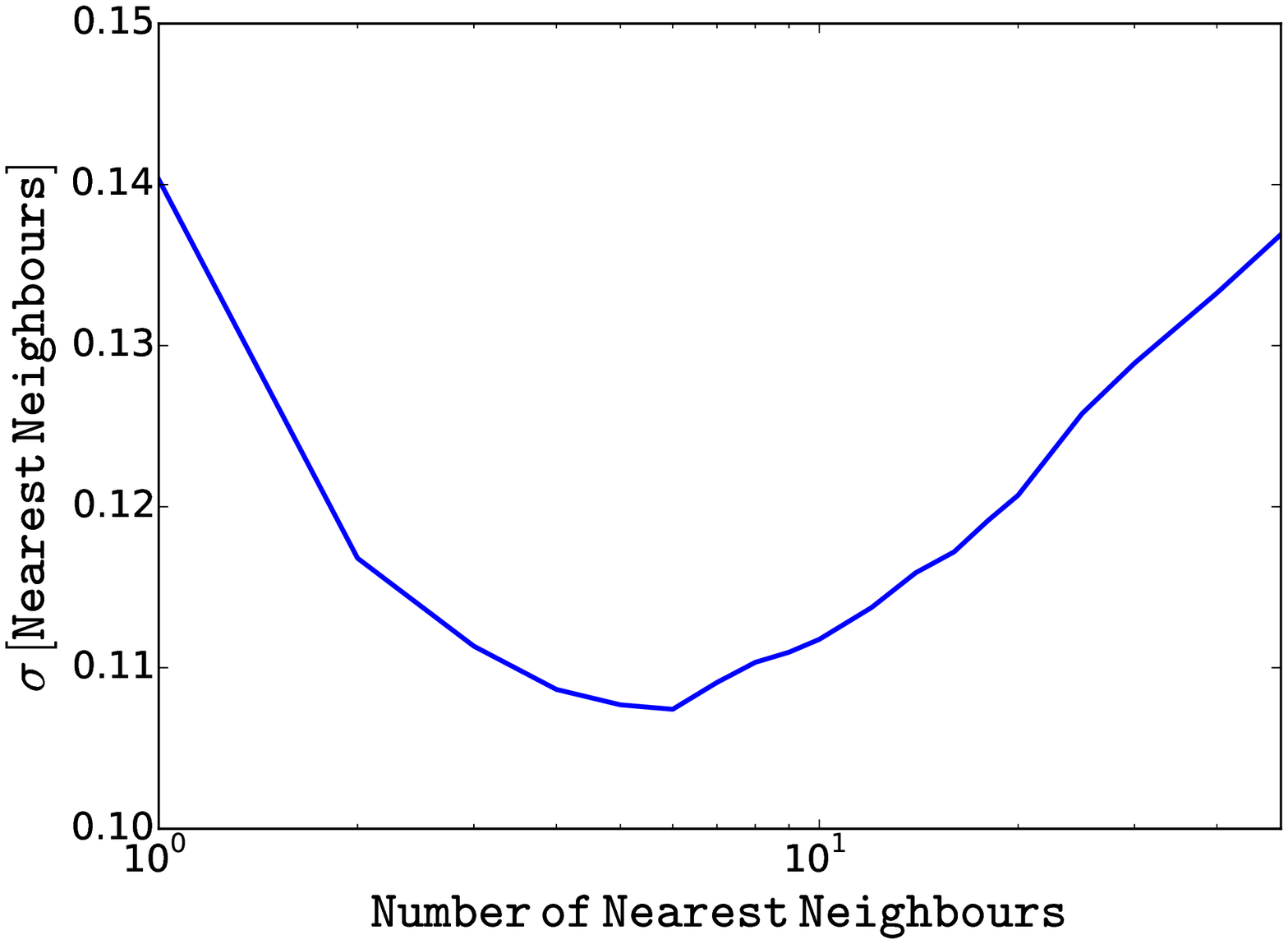}
	\includegraphics[width=0.45\textwidth]
		{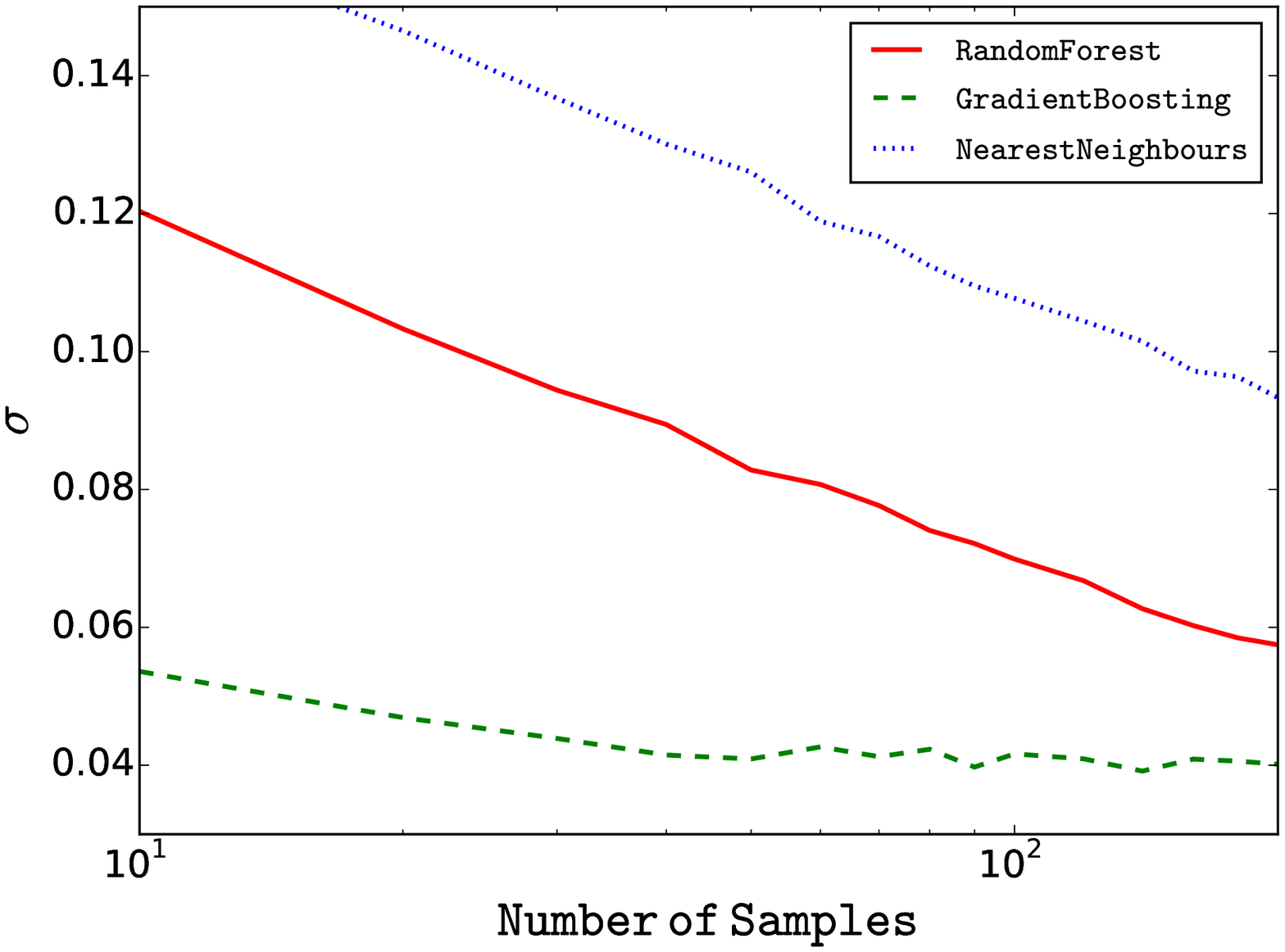}
	\caption{First three panels (in reading order) show the variation of the prediction
		accuracy (in terms of $\sigma$) with the prime internal parameter of
		each algorithm. Bottom right: Prediction accuracy versus number of
		samples in the training set for each of the algorithms (parameters set to optimal values)
}
	\label{fig:fraction}
\end{figure}

\section{Parameters optimization}\label{sec:params}

	In this section, we study the parameters of the data mining models to
	extract their optimal values. We used the {\bf R1000-1} dataset (as described
	in section \ref{sec:data}) for this exercise. This dataset contains 1 million samples, comprising 1000 cosmological models. We used a part
	of the dataset as the training set, on which we build the models, which we
	then evaluated on the remaining samples for validation. For each
	evaluation, we followed these steps :
	\begin{enumerate}
	\item Choose an algorithm and fix its parameters.
	\item Randomly select 10\% samples from the R1000-1 as the training set. Assuming perfect randomization, this corresponds to randomly selecting 100 sets of cosmological parameters.
	\item Train the given algorithm on the chosen training set.
	\item Predict the power spectra for the remaining 90\% samples (test set)
			based on the trained model.
	\item Compute the {\em relative errors} in estimation.
	\item Compute the {\em mean} and {\em standard deviation} for the relative
			errors.
	\end{enumerate}
	We did this for a wide range of parameters for each of the three algorithms.
	At each iteration, a new, random training set is generated. This, hopefully,
	tames the bias in the estimators, and over-fitting, if any.\\
	\\
	A word on notation, let $({\bf x},y)$ be a test sample, where ${\bf x}$ is a vector of
	length 8, containing the cosmological parameters, the redshift value, and
	the $k-$value, and $y$ is the corresponding value of the power spectrum.
	We use a model $M$, trained on chosen training data, to estimate the value
	of the power spectrum. Call the estimated value be $\hat{y}$. Then the
	{\em relative error} is given by the following expression:
	\begin{equation}
	\delta y_{rel} = \frac{y - \hat{y}}{y}
	\end{equation}
	For each test sample, we get a relative error value, and thus, we obtain an
	array of the size equal to the size of the test set. We compute {\em mean ($\mu$)}
	and {\em standard deviation ($\sigma$)} of this array using the following convention -
	\begin{equation*}
	\mu = \frac{1}{N} \sum_{i=1}^{N} \delta y_{rel,i}
	\end{equation*}
	\begin{equation}
	\sigma^2 = \frac{1}{N} \sum_{i=1}^{N} (\delta y_{rel,i} - \mu)^2
	\label{eqn:sigma}
	\end{equation}
	where $\delta y_{rel,i}$ is the relative error of the $i$th test example.
	\\
	We use $\sigma$ to quantify the prediction accuracy of the algorithm e.g. $\sigma=0.05$ implies that 68.3\% of the samples are predicted with accuracy better than 5\%.
	\\
\subsection{ML parameters}\label{sec:mlp}
	In this section, we optimize for the prime internal parameter of each algorithm by minimizing $\sigma$. As $\sigma$ quantifies the prediction accuracy, this can be used to identify the best configuration such that the algorithm performs ideally for a realistic choice of the training set.
	\begin{itemize}
	\item {\bf Random Forests:} As we described in the section (3.2), a random forests
	algorithm is built by taking an ensemble of decision trees. The most
	important parameter of a random forest based model is the number of trees in
	the ensemble. We varied the number of trees (herein referred to as {\em
	number of estimators}), and recorded the mean and the standard deviation of
	the relative errors following the aforementioned steps. This is shown in the upper left panel of the figure \ref{fig:fraction}. It can be seen on the picture that the optimal value for number
	of estimators is near 100.
	\item {\bf Gradient Boosting:} A gradient boosted tree algorithm, that we used
	here, is built by fitting an initial set of decision trees on the training set,
	iteratively modifying the estimates while reducing the error. The parameter
	of interest in this model is the number of initial decision trees chosen. We
	fixed all the other internal parameters of the model, and varied the number
	of trees. The mean and standard deviation of the relative errors of the test
	set were computed. The variation of the error standard deviation with the
	variation in the number of decision trees is shown in the upper right panel of the figure \ref{fig:fraction}. The
	optimal choice for initial trees is around 1000.

	\item {\bf K-Nearest Neighbours:} The K-Nearest Neighbours algorithm works by
	estimating the outcome for a test case by averaging the outcomes of its $K$
	neighbours among the training cases. Thus, it is expected that the number of
	neighbours plays an important role in how accurate the estimation is. We
	varied the number of neighbours, referred to as $K$ in our analysis. The bottom left panel of the
	figure \ref{fig:fraction} depicts the behaviour of the standard deviation of the relative
	errors. The graph looks like a parabola, which is very much expected from
	the description of the algorithm. Initially the number of neighbours is too
	small to be able to assist in prediction, e.g. $K=1$ implies that we base
	our prediction only on the nearest neighbour. And as we increase the number,
	we see a drop in the error. This is due to the fact that more neighbours are
	being used to compute the predicted value. As we keep on increasing $K$,
	after a threshold, a lot of neighbours are creeping into the set we are using
	for prediction, and a lot of these neighbours are not close enough to have a
	say in the final prediction. Thus the accuracy drops. We observe the optimal
	value of $K$ to be around 5.
	\end{itemize}

\begin{figure}
	\centering
	\includegraphics[width=1.1\textwidth]
				{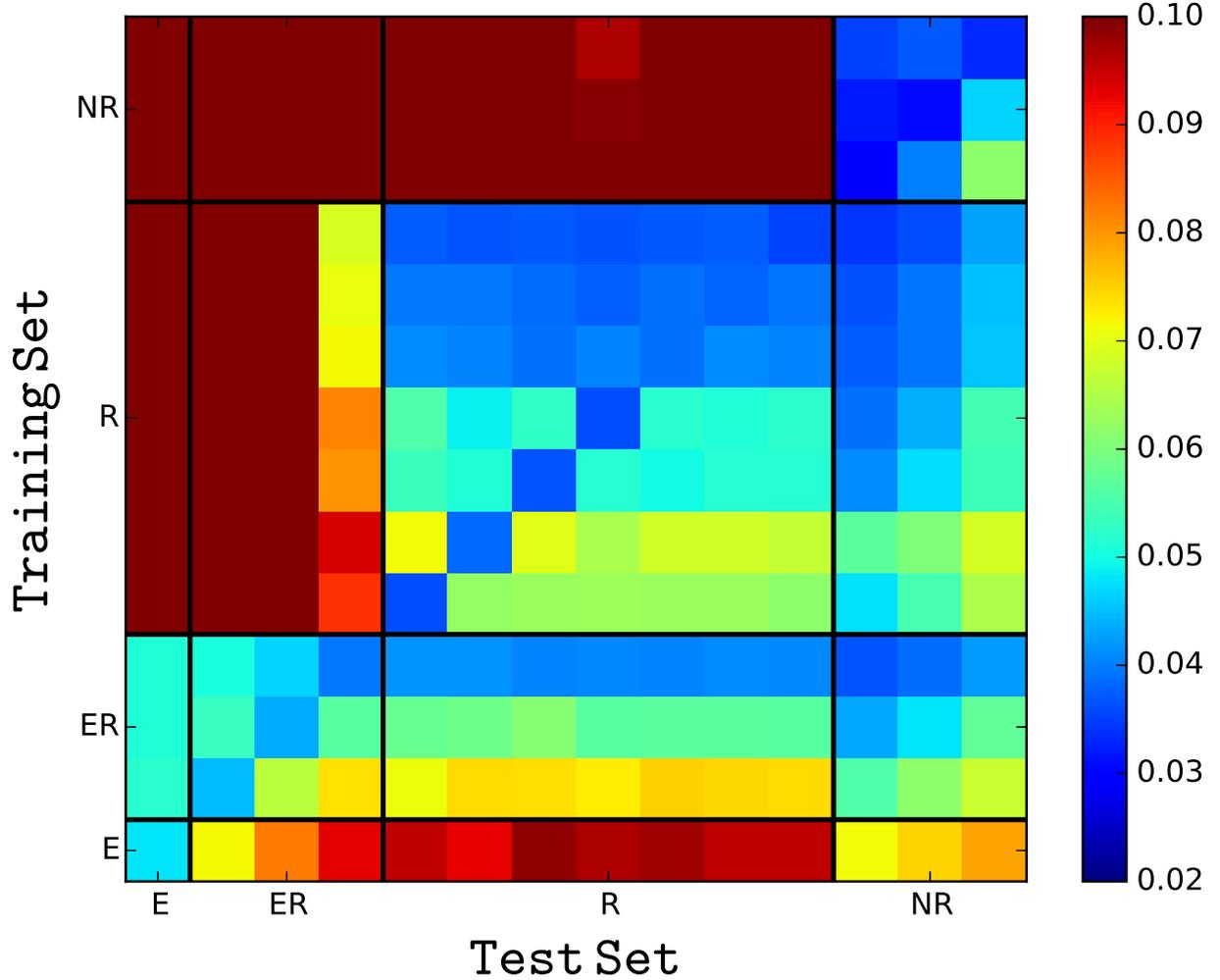}
	\caption{Accuracy (in terms of $\sigma$) in color codes for
			all the permutations of the datasets listed in table
			\ref{tbl:data}}.
	\label{fig:table}
\end{figure}

\subsection{Sample size}
We also studied the impact of the size of the training set on the accuracy of
the predictions, that we quantified in terms of $\sigma$ as defined in equation
\ref{eqn:sigma}.
We set the internal parameters of each algorithm to their optimal values as
compute in section \ref{sec:mlp}.
Bottom right panel of figure \ref{fig:fraction} shows the variation of $\sigma$
with the sample size varying between 10 to 200 cosmological models. As the
sample size increases, the predictions in each case become more accurate, which is to be expected.

The gain in accuracy for both KNN and RF is nearly by a factor of two when the size
of the training set is increased from 10 to 200, however, RF is performing much
better than KNN.

GBM, on the other hand, is doing much better than the other two. With a training set consisting of only 10 sets of
cosmological parameters, it can reach to an accuracy corresponding to $\sigma=0.06$.
If the size of the training set is increased, there is indeed a slight gain in the
accuracy, however $\sigma$ converges to about 0.04 after the size of the training set exceeds 50 cosmological models.
Therefore, if the number of models are limited (which is very much true in actual settings), GBM is the most promising
algorithm (of the three we tried) to get sufficiently good predictions over a large range of
cosmological models and redshifts.

\section{Predictions}\label{sec:predictions}

We used all datasets listed in table \ref{tbl:data}, and studied them as both
training set and test set for all possible permutations. Figure \ref{fig:table}
shows the resulting $\sigma$ in each case using GBM with 1000 {\it number of estimators}. Some
important features of the analysis are listed below:

\begin{itemize}
	\item All the training sets can be reproduced (by testing on the training set itself) with an average $\sigma~\sim~0.045$ (the diagonal of the figure \ref{fig:table}).
		It has a small scatter around it,
		depending upon the {\bf size} of the training set. The best recovered datasets
		are the ones with size 100, where $\sigma$ is close to 0.03.
	\item E38 dataset can reproduce itself with $\sigma\sim~0.05$.
		Testing a trained model on E38 yields a reasonably good accuracy ($\sigma < 0.1$) if the
		training set contains this set as well. If the E38 set is not included in the training set, we get $\sigma~>~0.1$.\\
		On the other hand, using only E38 as a training set, cannot predict any other dataset with $\sigma < 0.07$.\\ Thus, to correctly predict the power spectrum of samples in E38, we need to include E38 in the training set, and if only trained on E38, the model crashes for all other test sets.
	\item ER datasets appear to be the ideal training sets. They can predict
		E, R as well as NR datasets with $\sigma < .05$ if the training set comprises of more than 100 training cosmological models. If the size
		of the ER dataset is larger, the predictions become even better.
	\item R datasets cannot predict power spectrum for E or ER datasets with good accuracy, however,
		they predict all random datasets, both R and NR, with $\sigma \sim 0.04$ for 1000 cosmological models, and slightly larger $\sigma$ for smaller datasets.
	\item NR datasets are the worst training sets, except if the test
		set is also NR. In that case, the prediction is better than any other
		datasets permutation and reaches $\sigma~\sim~0.03$.
\end{itemize}
These observations have solid physical intuition. 
For example, E38 dataset comprises of the original 38 cosmological nodes of the cosmic
emulator, which have more accurate power spectrum compared to the other cosmological models evaluated through interpolation. Therefore, these 38 models
have only statistical noise, whereas other models also have a large systematic error. Hence, using E38 as a training set can only predict the power spectrum when the systematic errors are negligible. For all other test cases, like R and NR datasets, this is a highly biased training set.

All R datasets are compatible with each other, and only depend on the
sample size for its accuracy of predictions. This is due to the fact that all
R datasets comprise of similar systematics, and hence a single learning
algorithm can understand all at once.

The NR datasets can be predicted by any training set. This is because the NR
datasets have an 6-dimensional Gaussian distribution, and have a very high density
near the standard values (listed in table \ref{tbl:cosmo}). Due to the same
reason, NR-1 can be predicted much better than NR-3.

This analysis shows that if E38 models are important,
e.g. for a good compatibility with the cosmic emulator, ER datasets
make ideal training sets which yield $\sigma \sim 0.05 - 0.06$ with a realistic number of cosmological models ($\sim 100$). However, if the novelty of E38 data is not a big issue,
any random training set containing nearly 100 sets of cosmological parameters yields sufficient good estimates of power spectrum with $\sigma \sim 0.05$ up to
$k~\sim~10~h^{-1}~{\rm Mpc}$.


\section{Supervised Estimator for Matter Power Spectrum (SEMPS)}
\label{sec:semps}
\begin{figure}
	\centering
	\includegraphics[width=1.0\textwidth]
				{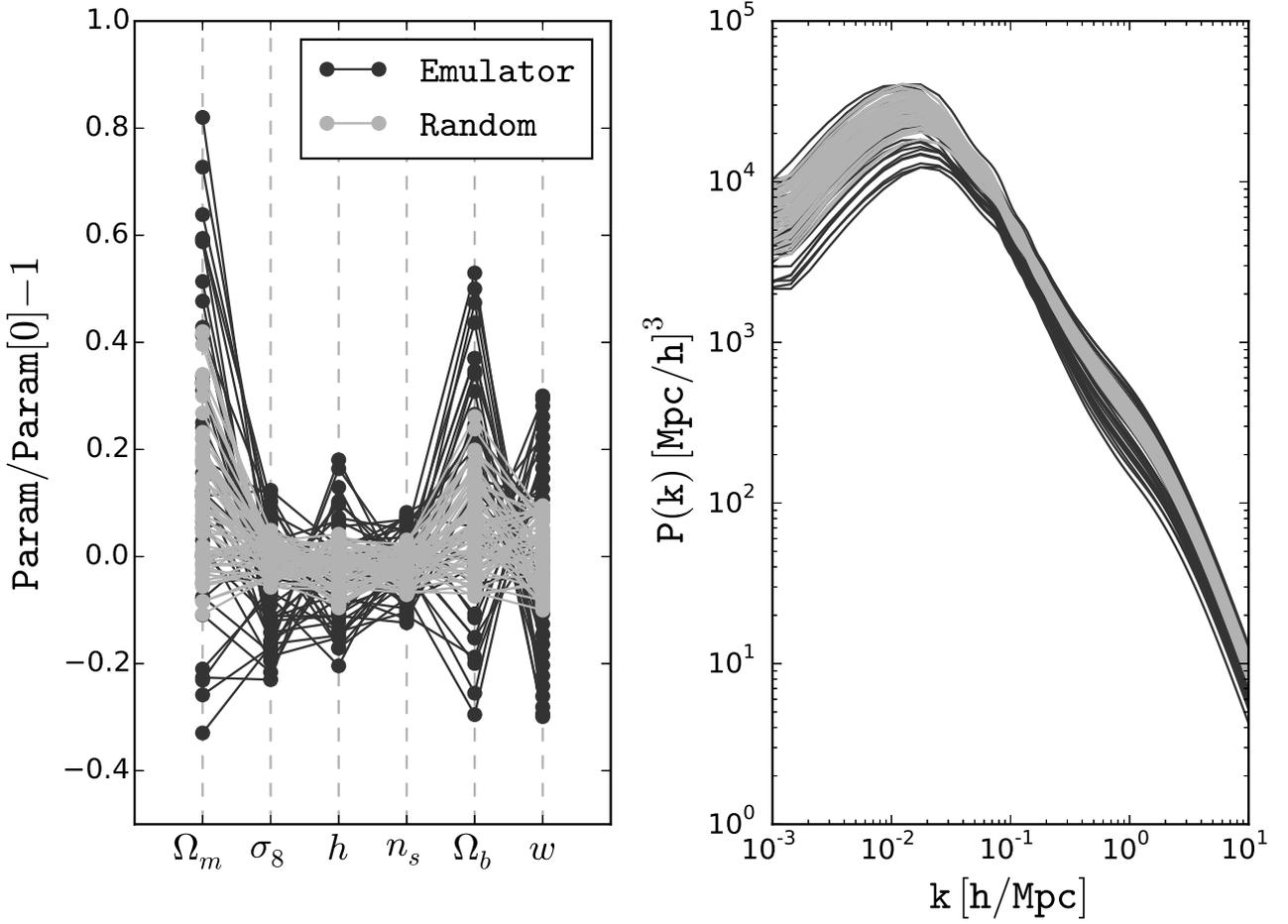}\\
	\caption{Left: 100 cosmological models of ER100 dataset. Right: the
			corresponding matter power spectra.}
	\label{fig:bb}
\end{figure}

\begin{figure}
	\centering
	\includegraphics[width=0.7\textwidth]
				{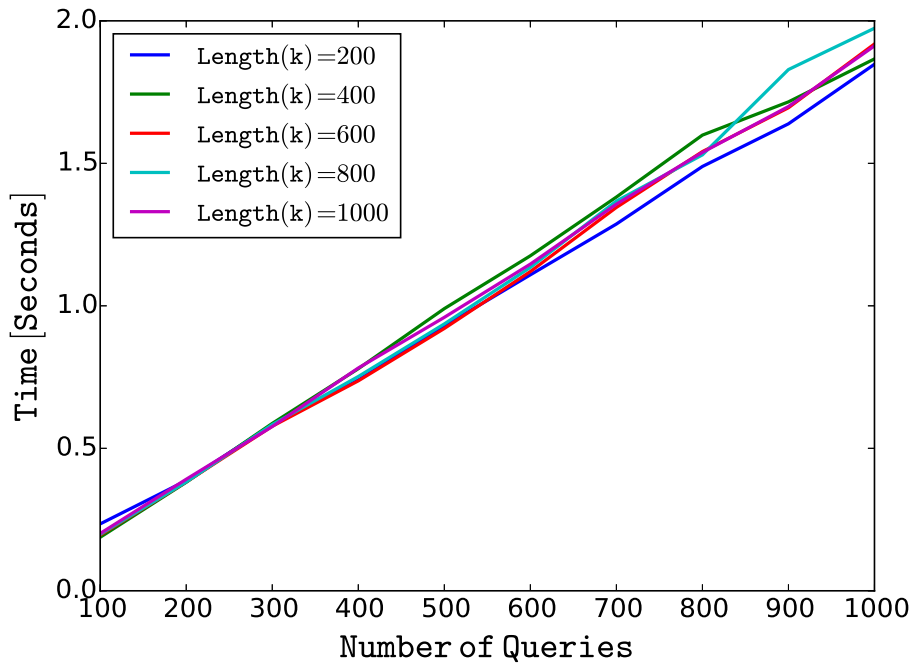}\\
	\caption{Computation time of the matter power spectrum using SEMPS
			as a function of the number of queries for different
			lengths of the $k$-array.}
	\label{fig:time}
\end{figure}

\begin{figure}
	\centering
	\includegraphics[width=0.45\textwidth]
				{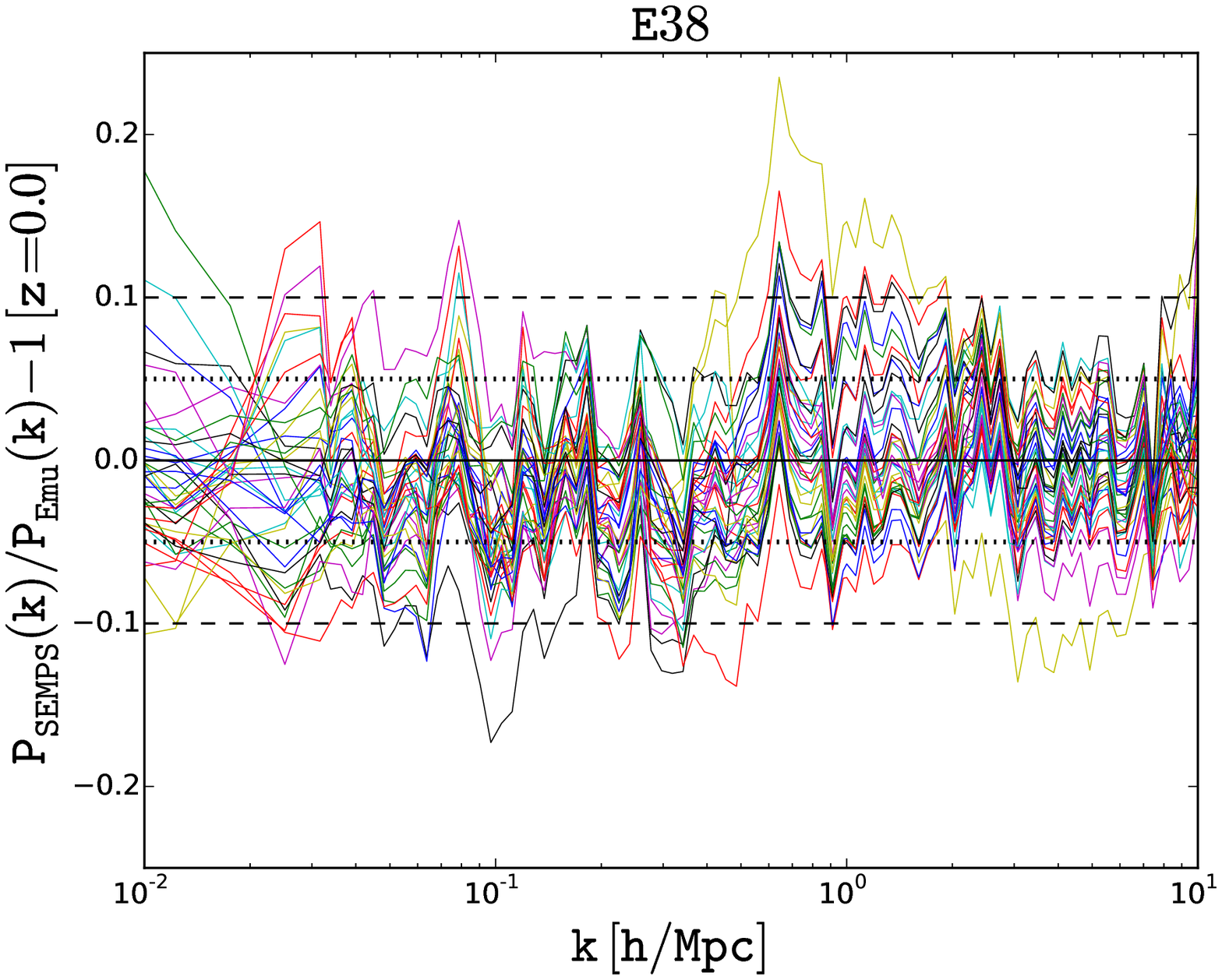}
	\includegraphics[width=0.45\textwidth]
				{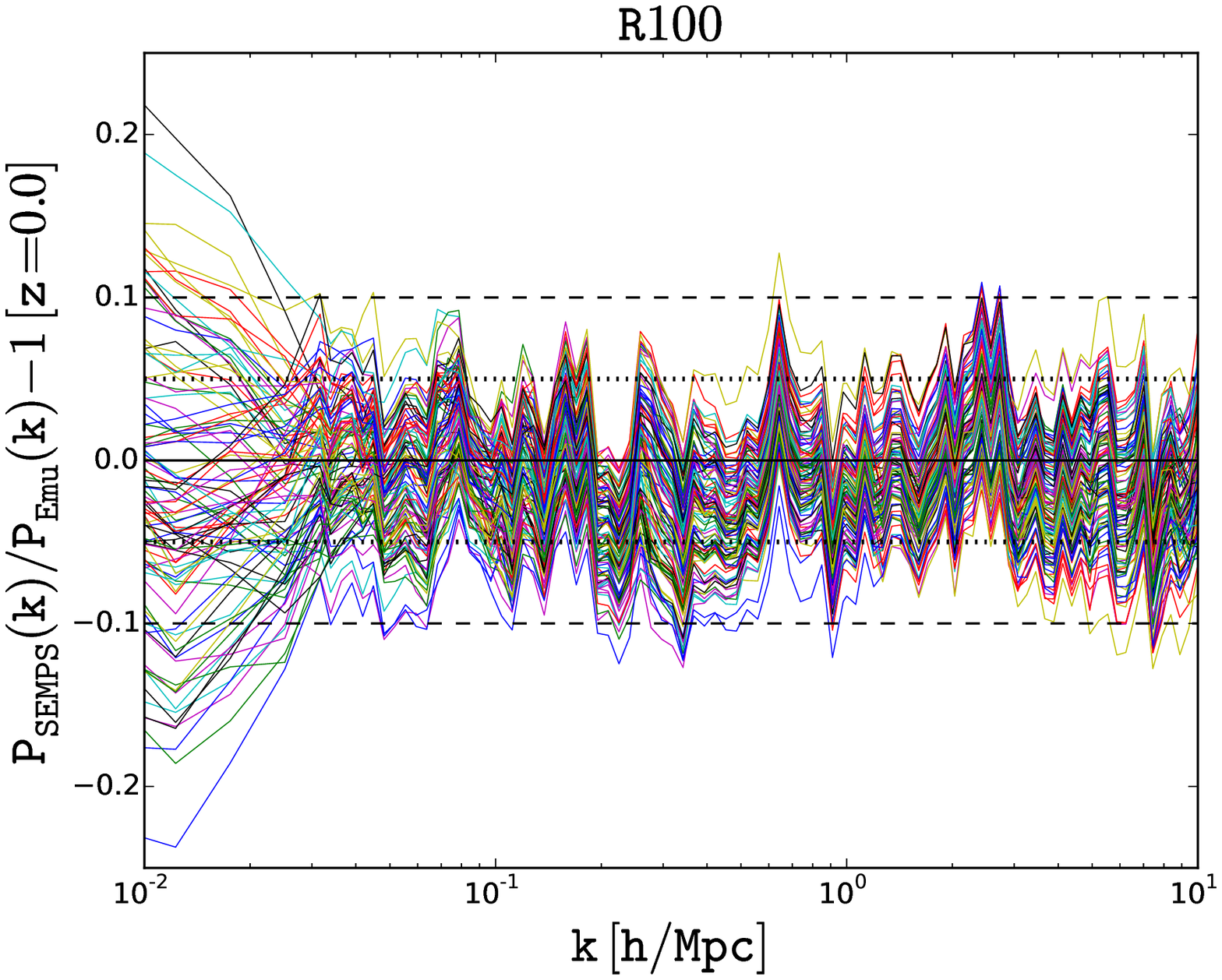}\\
	\includegraphics[width=0.45\textwidth]
				{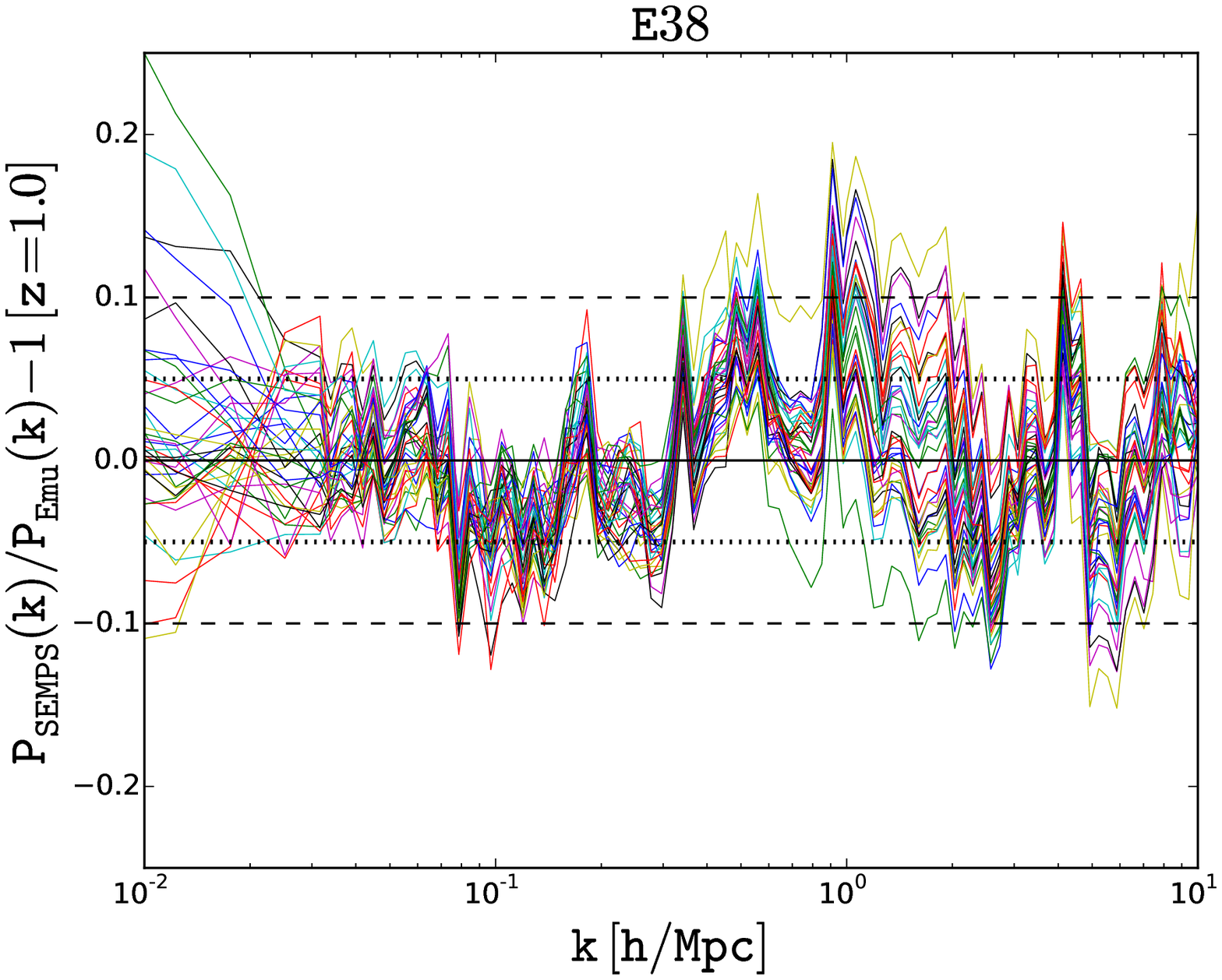}
	\includegraphics[width=0.45\textwidth]
				{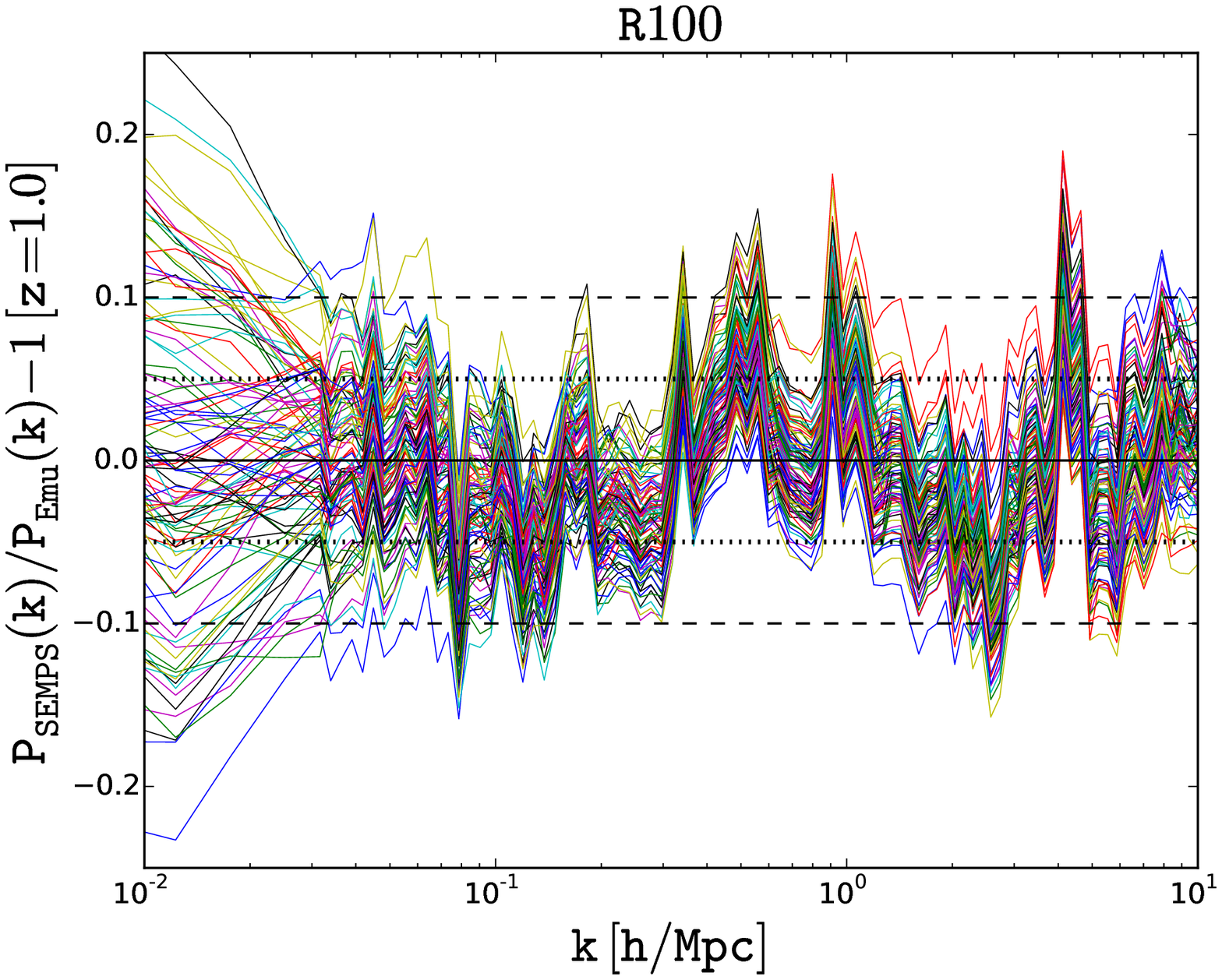}\\
	\includegraphics[width=0.45\textwidth]
				{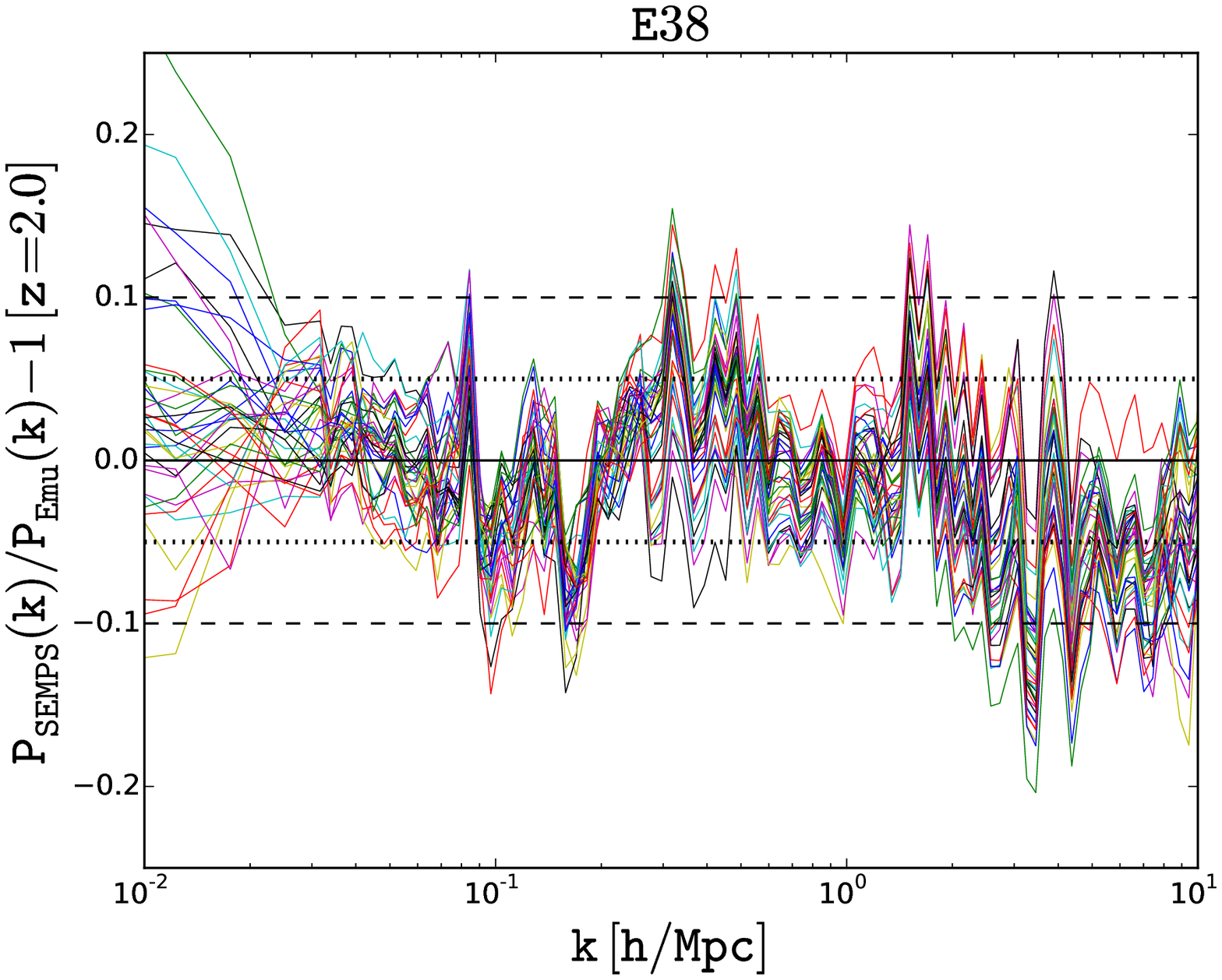}
	\includegraphics[width=0.45\textwidth]
				{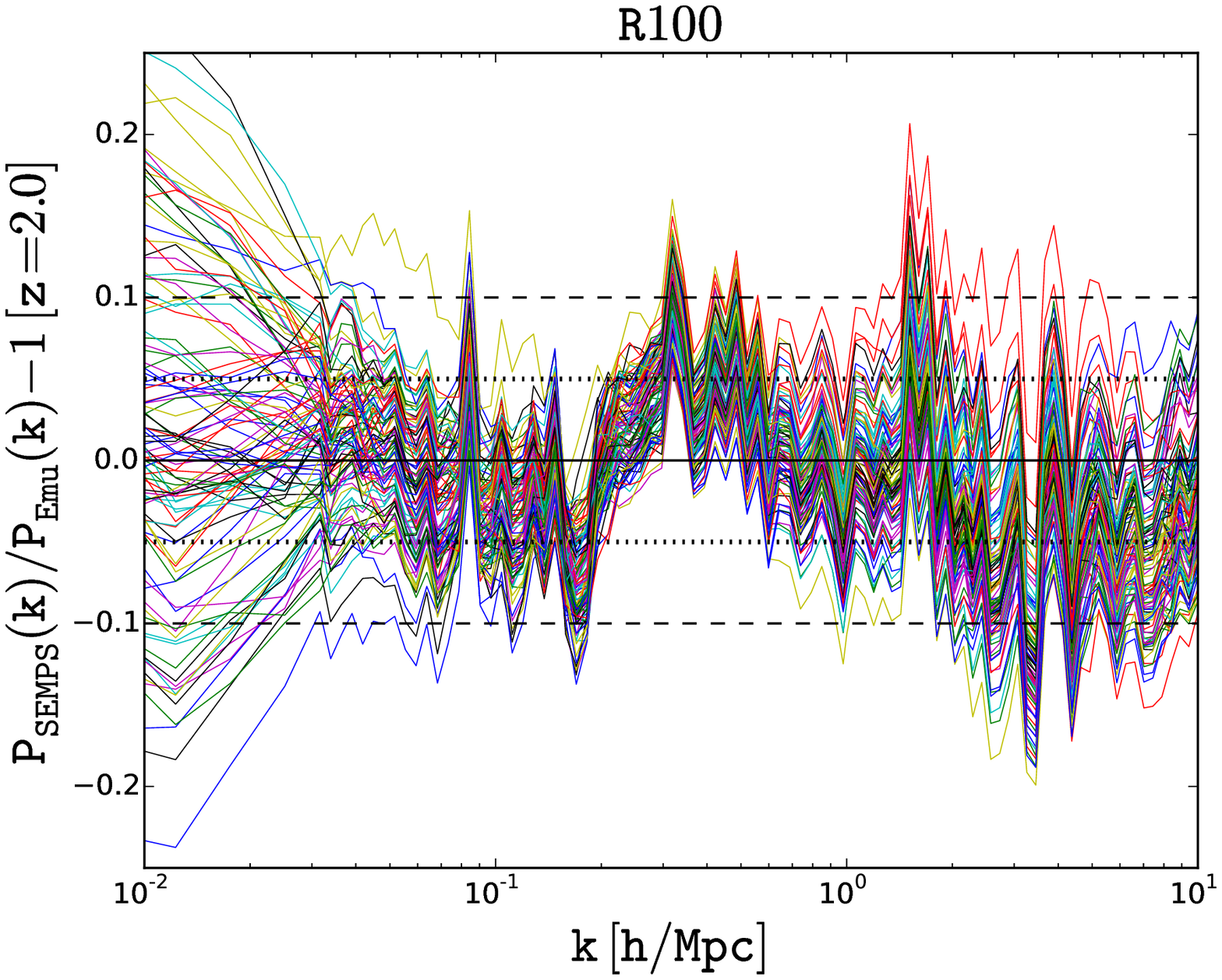}\\
	\caption{Estimates of the matter power spectrum from SEMPS
			and the emulator for two datasets: E38 (left column) and
			R100 (right column) for three different redshifts in different
			rows.}
	\label{fig:compare}
\end{figure}

\begin{figure}
	\centering
	\includegraphics[width=0.9\textwidth]
				{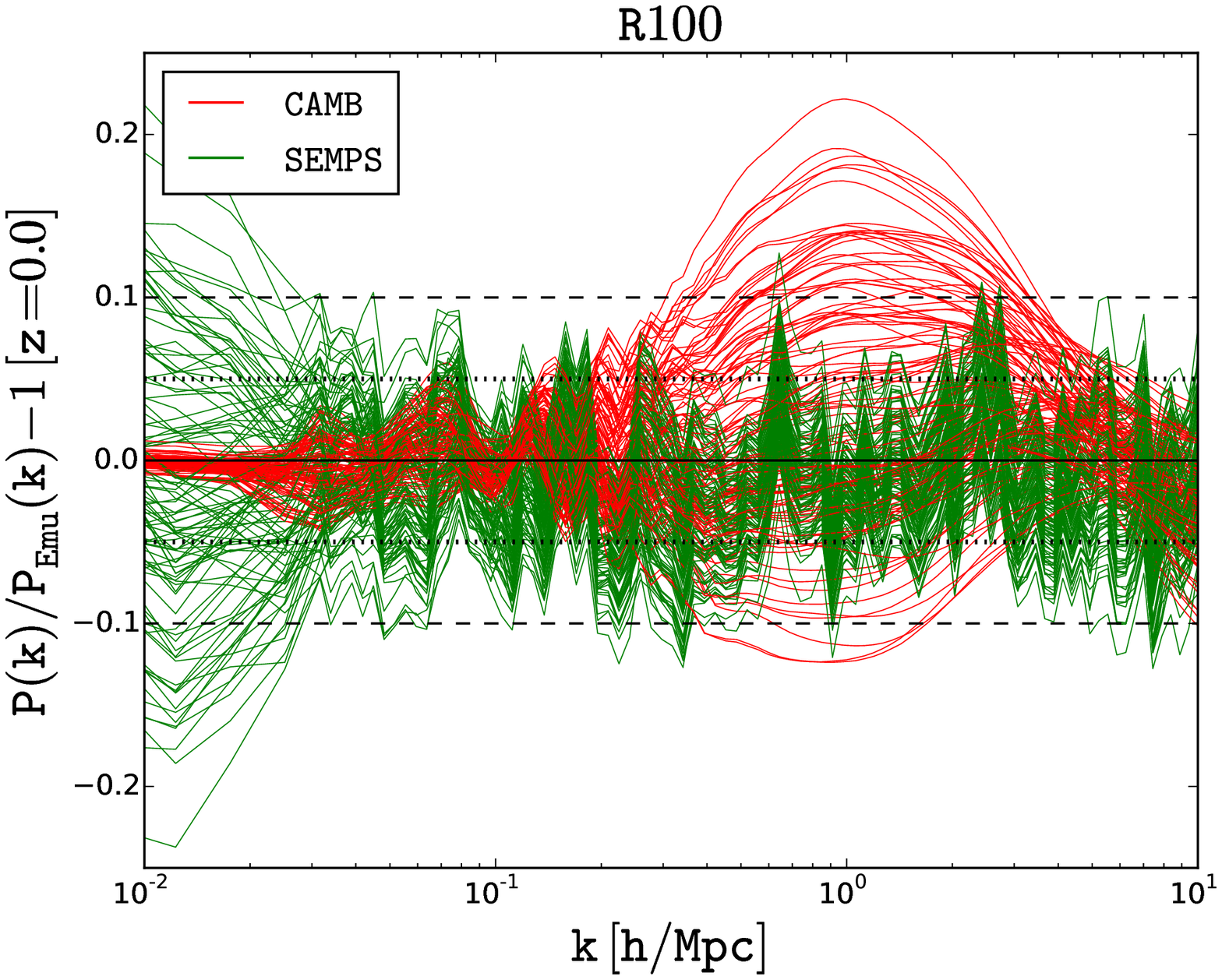}\\
	\caption{Comparision of matter power specturm estimates from
			SEMPS and CAMB with those from the emulator on R100 dataset at
			redshift zero.}
	\label{fig:camb}
\end{figure}

With the collective understanding from previous sections,  we build a blackbox
called SEMPS --- a
model trained on ER100 using Gradient Boosting with 1000 number of estimators
--- and provide it as a supplement to the paper.
We performed a few tests to show its productivity and usefulness. Figure
\ref{fig:bb} shows the cosmological models and the corresponding power spectra
of ER100 dataset at redshift zero.

\subsection{Prediction time}

We studied the efficiency of SEMPS
on a variable length of ${\mathbf k}$. Figure \ref{fig:time} shows the prediction time
versus the number of queries made. It is evident that the length of the ${\mathbf k}$
vector does not matter for the prediction, however, the prediction time goes
linearly with number of queries. Therefore, in order to evaluate an observable
which is an integrated quantity over the matter power spectrum, for example weak
lensing shear power spectrum, 200-500 queries are needed, and hence it can be
evaluated in less than one second. To minimize the interpolation errors, power spectrum can be evaluated on a finer $k$-array, while the prediction pipeline stays equally efficient.

\subsection{Comparison with cosmic emulator and CAMB}

In this section, we explicitly show the comparison between the matter power
spectrum estimation of the SEMPS and the true emulator values.
Figure \ref{fig:compare}  shows the comparison for the
E38 dataset (left column), and R100 dataset (right column) at three different
redshifts (different rows).
In both cases, the predictions are better than 10$\%$ accurate
up to two sigma, and 5$\%$ accurate up to one sigma error bars.

An important aspect is that the predictions of the matter power spectrum in both
cases are statistically similar, although both are very distinct datasets.
Therefore, SEMPS provides an unbiased estimator. Also, the errors are
distributed equally for all $k$-ranges, and hence an integrated quantity can
average out the noise leaving the observable independent of any systematics.

In figure \ref{fig:camb}, we compare the matter power spectrum prediction from
both SEMPS and CAMB \cite{} to the emulator output. For small values of $k$,
CAMB seems to match better with Emulator compared as to the SEMPS predictions,
however, in this regime the cosmic variance is large, and the accuracy is not
required. For $k>0.1$, where the cosmic variance drops significantly, SEMPS
estimations happen to match better than CAMB predictions.

\subsection{Software package}
An open source python software package for SEMPS is available for use, and
further developments. It can be downloaded from
\url{http://www.ics.uzh.ch/~irshad/semps/} as a $\mathtt{tar}$ file which contains
the code for SEMPS, and an illustrative example file.


\section{Discussion}\label{sec:discussion}

In this article, we reviewed three machine learning algorithms --- RF, GBM, and
KNN --- and analysed their usefulness in predicting the matter power spectrum.
We also computed the optimal ranges of prime internal parameters for each
algorithm which maximize the accuracy for a training set. We found that for RF
and GBM, where the parameter under consideration is the number of estimators
which controls the complexity of the tree models, it is 100 and 1000
respectively. For KNN, the prime parameter, number of nearest neighbours, has an
optimal value of 5. Fixing the optimal values for each prime parameter,
we studied the importance of the size of the training set, and found the best
behaviour being reported by GBM for all training sets containing 50-100 sets of
cosmological parameters (see figure \ref{fig:fraction}).

Inspired by our analysis, we chose GBM with 1000 estimators as the prime
algorithm, and studied many different datasets, of different sizes and nature,
for their role as training and test sets. We found that the
self- predictability of any set is 3-4 $\%$, whereas the cross predictability
depends on many characteristics (see figure \ref{fig:table}).
We found that the data generated by the cosmic emulator, E38 dataset, is highly
biased in our analysis because it does not share the
similar systematics that exist in all other randomly generated datasets, and in
order to predict E38 with a good accuracy, those cosmological models have to be
included in the training set. However, a random dataset containing 50-100
cosmological models is good enough to estimate a matter power spectrum better
than 10$\%$ up to $k~\sim~10~h^{-1}~{\rm Mpc}$ for a wide range of cosmological
models and redshifts.

We developed an estimator, SEMPS, for the matter power spectrum, which can be
used as a blackbox, and is intended to be used this way.
It is trained on ER100 (see figure \ref{fig:bb}) dataset which comprises of a total
of 100 cosmological models, 38 of which are cosmic emulator's original nodes,
and remaining 62 are random set of cosmological parameters. It was
trained using GBM algorithm with a total of 1000 estimators.
We analysed the prediction time of the blackbox, and showed explicitly that up
to 500 matter power spectra can be computed, for any length of the $k$-array,
in less than one second (see figure \ref{fig:time}).
We compared the matter power spectrum predicted from SEMPS
to that of the cosmic emulator, and found an agreement of $\sim 5\%$ up to
one-sigma, and $\sim 10\%$ up to two-sigma for all datasets considered
(see figure \ref{fig:compare}).
We also compared the estimations of SEMPS to that of the widely used CAMB,
(Halofit). SEMPS is found to be in a better agreement with the emulator
values than CAMB (see figure \ref{fig:camb}).

Our estimations show wiggles for all cosmological models and redshifts.
This is due to the reason that it does not have any functional form, but a tree
structure that is discrete in nature. On the more optimistic side, these wiggles
are distributed uniformly throughout the $k$-range, cosmological models and
redshifts. Therefore it is possible to average them out by binning in $k$, large
enough to include the wiggles. Due to the large number of estimators, the scale
of these wiggles are very small, and hence a mean matter power spectrum can be
derived from SEMPS. However, a major drawback of this method is
that it will erase any small scale features, like Baryon Acoustic Oscillations
at $k\sim 0.1 h^{-1}{\rm Mpc}$. Therefore, particularly for the applications
which do not include BAO feature, this method can supply a smooth matter power
spectrum estimator, and not otherwise.

\subsection{Applications}
The accuracy achieved with SEMPS, and its computation being instantaneous, make
it an ideal tool for the following cosmological applications.

\begin{itemize}
    \item Cosmological observables: As matter power spectrum underlies many
        cosmological observables -- for example galaxy clustering, weak lensing
        shear power spectrum etc. -- an accurate and unbiased estimation of the
        matter power spectrum is needed. Particularly, for the weak lensing
        applications, SEMPS predictions can be very useful as the weak
        lensing observable is an integrated quantity of the matter power
        spectrum over the redshift, it can average out any noise that exist
        in SEMPS estimates. Also, for this purpose the BAO feature is not very
        important, so the bin-smoothing of the matter power spectrum can be very
        useful.

    \item Building emulators: As shown in the analysis that SEMPS
        predictions are NOT largely biased to its training set, and is
        predicting the matter power spectrum with noise nearly independent of
        any cosmological model, it makes it an unbiased estimator. Such
        estimators can certainly be employed to make emulators for other
        cosmological observables like galaxy power spectrum, redshift space
        distortions, two-point correlation function etc.

    \item Visualisation: due to the operational speed of SEMPS,
        it is possible to build interactive tools to visualise the sensitivity
        of the matter power spectrum to the cosmological parameters, and
        redshifts. This can be very useful to explore degeneracies in the
        cosmological parameters, or simply for educational purposes. 
\end{itemize}

\subsection{Prospects}
SEMPS is built on matter power spectra derived from 100
cosmological models computed with the cosmic emulator. The cosmic emulator
itself is accurate only up to $5\%$ at its original cosmological nodes, and
lesser so for other models. Therefore, the current blackbox is sharing an
underlying systematic and interpolation error that were induced by its own
training set.
To build an improved estimator, a set of 50--100 cosmological simulations can be
run, and a similar model can be trained on thus computed matter power spectra.
Such an estimator will be independent of any systematic or interpolation errors.

To run such simulations, few million CPU hours would be sufficient.
It is certainly possible with the current computational resources that an
emulator can easily be built which is light-weight, fast, and accurate.
It is also possible to build a flexible emulator system such that new data can
be added when available.
With current advancements in the computational resources, large cosmological
simulations can be run efficiently and inexpensively. In fact, such simulations
are being run very often in the cosmological community. Most of these
simulations are being run for standard cosmological models derived from recent
constraints, for example Planck \citep{2015arXiv150201589P}.
Therefore, if power spectra
derived from such simulations are added into the training set, the distribution
of the cosmological parameters approaches from a uniform to a normal
distribution. This will further improve the accuracy of the matter power
spectrum estimates in the range overlapping with priors from current data.

\clearpage

\bibliographystyle{mnrasfile}
\def\apj{ApJ}
\def\fcp{fcp}
\def\apjl{ApJL}
\def\aj{AJ}
\def\mnras{MNRAS}
\def\aap{A\&A}
\def\nat{Nature}
\def\pasj{PASJ}
\def\prd{PRD}
\def\physrep{Physics Reports}
\def\jcap{JCAP}
\bibliography{ms.bib}
\newpage

\end{document}